\documentclass[12pt]{article}
\usepackage{amsmath}

\usepackage{graphicx}
\usepackage{verbatim}
\usepackage[utf8]{inputenc}
\usepackage[T1]{fontenc}
\usepackage{changepage}
\begin{document}

\title{\LARGE\bf A Statistical Framework for Single Subject Design with an Application in Post-stroke Rehabilitation}

\author{\large Ying Lu\thanks{Center for the Promotion of Research Involving Innovative Statistical Methodology, Steinhardt School of Culture, Education and Human Development, New York
University, New York, USA}, \quad  \large Marc Scott$^*$ %\thanks{Center for the Promotion of Research Involving Innovative Statistical Methodology, Steinhardt School of Culture, Education and Human Development, New York University, New York, USA},
\quad and \quad
{\large Preeti Raghavan }\thanks{Department of Rehabilitation Medicine, New York University School of Medicine, New York, USA}
}
%The first and third authors thanks the
%NIH xx grant support \vspace{-0.8cm}

\maketitle

\begin{abstract}
This paper proposes a practical yet novel solution to a longstanding statistical testing problem regarding single subject design. In particular, we aim to resolve an important clinical question: does a new patient behave the same as one from a healthy population? This question cannot be answered using the traditional single subject design when only test subject information is used, nor can it be satisfactorily resolved by comparing a single-subject’s data with the mean value of a healthy population without proper assessment of the impact of between and within subject variability. Here, we use Bayesian posterior predictive draws based on a training set of healthy subjects to generate a template null distribution of the statistic of interest to test whether the test subject belongs to the healthy population. This method also provides an estimate of the error rate associated with the decision and provides a confidence interval for the point estimate of interest. Taken together, this information will enable clinicians to conduct evidence-based clinical decision making by directly comparing the observed measures with a pre-calculated null distribution for such measures. Simulation studies show that the proposed test performs satisfactorily under controlled conditions.

\end{abstract}

\section{Introduction}
Making an inference regarding a single subject is an important goal in clinical and applied settings and in health and behavioral research (for examples,\cite{Kazdin1982}, \cite{Kenndy2004} and \cite{GastLedford2014}). Here, one is often interested in assessing an individual subject's outcome across different behavioral conditions. When only a single data point is available for each condition, one can only make a visual judgment regarding the direction and magnitude of the change. Kazdin~\cite{Kazdin1982} recommended performing repeated trials under the same condition to reduce the impact of within subject variability using a repeated measures design such as the ABAB type, where A and B each refers to different conditions. In this case, one can perform a within-subject statistical test to evaluate the change using a randomization based test (\cite{Onghena1992} and \cite{Onghena2005}). However the classic single subject design and method of analysis does not allow clinicians and researchers to compare the test subject with a reference population and assess the impact of between subject variability on the decision.

The primary goal of this paper is to demonstrate a method to make an inference regarding the behavior of any single subject from a group of test subjects given an available training set of subjects whose status is known. Instead of making an inference about the average behavior of the test set as a group, we are interested in assessing the status of test subjects individually or in small groups and seek to answer questions such as: {\it Does the test subject behave the same as someone in the healthy population as characterized by the subjects in the training set? } Note that in this setup, the sample size of the training set may be very small.  The training and the test sets both have repeated measures design, but the number of trials may not be the same.  Moreover the experimental conditions for the test may only be a subset of the training set.

In this paper we propose a novel statistical framework for testing the above question in the context of a single subject experiment, given a small amount of training data. A simple test statistic based on sample mean difference between conditions for the test subject is compared to a template distribution as a surrogate for the true sampling distribution of the mean difference under the null hypothesis. This template distribution is generated based on Bayesian posterior predictive draws using the training data set and the single-subject design. In the remainder of the paper, we first introduce a motivating example from post-stroke hand rehabilitation where a change in the fingertip grip and load forces and the rate of change are key variables for assessing the quality of motor control in a grasping task. We then discuss several standard statistical models that attempt to answer a research question in rehabilitation. Next, we outline the proposed method for testing the performance of a single subject, followed by a simulation study that examines the proposed method in comparison to more traditional approaches. The power of hypothesis testing with different single subject designs is also studied using simulations.  Lastly, we analyze a real data set of change in fingertip forces when grasping and lifting a device in the context of hand rehabilitation to demonstrate the utility of the method for clinicians and other practitioners.

\section{An Example from Fingertip Force Regulation}
 An objective assessment of hand function is one of the most sensitive tests of neurologic dysfunction. Precision grasp is important for a number of daily activities such as grasping a cup of coffee, or picking up an egg. Johansson et al. initiated the examination of fingertip force coordination during grasping [see \cite{Johansson1998} for a review], and their results have been used as a model to study sensorimotor integration for more than 30 years. This model has been found to be effective in detecting impairment of fine motor control in various patient populations (see for examples, \cite{Fellows1998}\cite{Reilmann2013}\cite{Gordon1999}\cite{Gordon1997}\cite{Schwarz2001}). Efficient fingertip force coordination requires the ability to predict the optimal force when lifting an object, such as a cup of coffee. In healthy individuals, it has been found that after just one or two practice lifts, the rate of change of load force is faster for a heavier object than for a lighter object (\cite{Johansson1988b}\cite{Flanagan2000}). Recently, Lu et al.\ (\cite{Lu2015}) showed that after one practice trial, the peak rate of change of load force increases proportionally as the weight of the object being lifted increases.

However, predictive control of fingertip forces and movements is often impaired in patients with brain injury due to stroke (\cite{Hermsdorfer2003}\cite{Raghavan2010b}\cite{Raghavan2006}). The assessment and restoration of predictive control of fingertip forces has implications for diagnosis, prognosis and treatment of neurologic conditions, such as stroke, multiple sclerosis, Parkinson’s disease, cerebral palsy etc.  A specially constructed device instrumented with force sensors can  readily measure the grip and load forces during a grasp and lift task which involves lifting different weights (\cite{Lu2015}\cite{Raghavan2006} ), and it can be used as a convenient clinical metric to assess hand performance.

According to Lu et al.\ (\cite{Lu2015}), the logarithm of the peak Load Force Rate (PLFR) increases linearly with the object's weight among healthy subjects even before the object is lifted (see an illustration in Figure 1). Mathematically, we can express this as:
\begin{equation} \label{eq:basic_eqn}
log(\mbox{PLFR}_i) = \alpha_i + \beta \mbox{WEIGHT}_{it} + \epsilon_{it},
\end{equation}
where individual $i$ is lifting weight $\mbox{WEIGHT}_{it}$ on trial $t$, and $\epsilon_{it}$ is the idiosyncratic error.  The terms $\alpha_i$ and $\beta$ reflect individual-level baseline force and population-level (common) effects for different weights, respectively. Based on a sample of 10 healthy subjects, the scaling factor is found to be 1.4 Newton/ms per 1000 grams weight increase. Moreover, although individual subjects may have different rates of increase in the load forces prior to object lift, the manner in which PLFR {\it scales up} as a function of weight is fairly constant. In other words, in the above linear model, each subject could have his/her own intercept ($\alpha_i$), but all the subjects shared a common slope (the scaling factor $\beta$).

%{\it A new device...mention the availability of the PLFR measure in clinical setup}

Under the framework of model (\ref{eq:basic_eqn}), assessment of predictive control of fingertip forces can then be formulated into the following hypothesis testing problem.
\begin{description}
\item[H0: Patient has normal predictive control] The PLFR of the test subject increases as the weight of the object increases in the same way as in the healthy population. $\beta^{test}=\beta^{pop}$.
\item[Ha: Patient does not have normal predictive control] The test subject fails to adjust the load force rate due to weight changes, hence PLFR does not increase the same way as in the healthy population,  $\beta^{test}< \beta^{pop}$
\end{description}

\begin{figure}[ht]
\begin{center}
\includegraphics[width=4in]{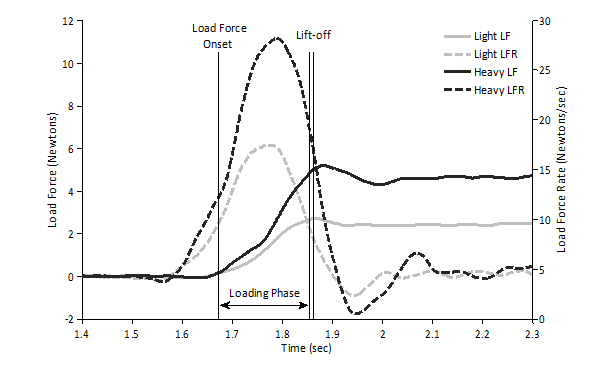}
\caption{Rate of Change in Load Force when Lifting Objects of Different Weights: The load force profiles of a subject grasping the grip device of two different weights with precision grip and lifting it with the dominant hand are shown in solid line. The derived force rate curves are shown in dashed line. Black indicates heavier object and grey indicates lighter objects. The peak Load Force Rate is defined as the highest point in the force rate profile.}
\end{center}
\end{figure}\label{LFR}

To test this hypothesis, we need to estimate the benchmark value for the scaling factor in the healthy population, $\beta^{pop}$, and the scaling factor for the test subject, $\beta^{test}$. Moreover, since the PLFR is a behavioral measure, there is substantial trial to trial variability when the same subject is lifting the same weights over multiple trials, and there is substantial between subject variability due to individual behavioral idiosyncrasy (\cite{Lu2015}). Hence a good statistical test should take into account the uncertainty introduced by both between and within subject variability. In the next section, we will introduce a typical dataset and comment on several approaches to assessing a single subject using existing statistical modelling techniques.
\section{Data and Existing Approaches}\label{classic_methods}
\subsection{The Data}\label{data}
First we describe the data that is typically obtained during precision grasp experiment. We have available a small training data set of 10 healthy subjects, lifting 10 weights ranging from 250 to 700 grams, 50 grams apart \cite{Lu2015}. Each subject lifts each weight over 7 consecutive trials, where the first trial is a learning trial as the new weight is presented in a random order and unknown to the subject. After the first trial, healthy subjects are capable of predicting the load forces and the load force rates for the given weight before the object is lifted in the subsequent 6 trials \cite{Lu2015}.  While the number of subjects is small, this data set is reasonably large given the number of conditions and repetitions, allowing for precise estimation of underlying physiological features and their variation.

The test set consists of some stroke patients.  Due to time and physical limitations, each test subject only lifts two to three different weights and for each weight, the test subject performs one practice lift in order to learn the weight of the object, then repeats for fewer number trials.

\subsection{A Natural Estimator}
In a clinical setup, the most straight-forward way to estimate the scaling factor of the test subject is to simply take the difference in the peak load force rate measured at different weights, averaged over multiple trials.  If the subject only lifts two weights, a naive sample mean based estimator of the scaling factor is,
\begin{eqnarray*}
\bar{\beta}_i  =  \frac{\sum_{t=1}^T(y_{i2t}-y_{i1t})}{T(w_2-w_1)}
\end{eqnarray*}
where $y_{i1t}$ is the PLFR measure for subject $i$ lifting weight one of $w_1$ grams at the $t$th trial ($t=1,\ldots,T$), and $y_{i2t}$ is the corresponding measure when the subject lifts weight two of $w_2$ grams at the $t$th trial.

Sometimes, the test subject can be instructed to lift three or more weights. In this case, the scaling factor $\beta$ can be estimated by averaging of the differences in PLFR between pairs of weights divided by the weight difference per pair. We propose the following estimator that averages all possible pairings, equally weighted, but other approaches are possible.
\begin{eqnarray}
\bar{\beta}_i  =  \frac{1}{\sum_{k=1}^{J-1}\vert\mathcal{I}_k\vert}
\sum_{k=1}^{J-1}\sum_{(j,j')\in\mathcal{I}_k} \sum_{t=1}^T \frac{(y_{ijt}-y_{ij't})}{T(w_j-w_{j'})} \label{eq:nparm0}
\end{eqnarray}
where $\mathcal{I}_k=\{(j,j'): j-j'=k\}$, and $w_j$ or $w_{j'}$ are weights associated with the corresponding indices.
%old
%, in which case, the PLFR can be estimated by averaging of the differences in PLFR between adjacent pair of weight then divided by the weight difference per pair. Putting it in a general formula assuming $J$ weights with equal distance $d_w=w_2-w_1$,
%\begin{eqnarray}
%\bar{\beta}_i  =  \frac{\sum_{j=2}^J \sum_{t=1}^T (y_{ijt}-y_{i(j-1)t})}{(J-1)T d_w} \label{eq:nparm0}
%\end{eqnarray}
The proposed naive estimator is fairly easy to compute. It also has several useful properties. First, it is an unbiased estimator. Second, we note that for up to four weights, this naive estimator is equivalent to fitting the slope via least squares estimation using the mean of repeated trials, $\bar{Y}_{ij\cdot}$, as (combined) observations. For least squares estimators, if the range of the weights is fixed, then an equal-distance weight design is most efficient.\footnote{The variance of the OLS estimator is proportional to the inverse of the (co)variance of the weight design $W$, $(W^\top W)^{-1}$. If the range of $W$ is fixed, the equal-distance design of the weights has the largest variance among other weight distributions. } Third, the variance of this estimator is inversely proportional to the distance between each pair of weights. Hence whenever feasible, one should choose weights that are further apart. Lastly, we note that just as with the ordinary least squares(OLS) estimator, for equal-distance weight designs, using two weights or three weights over the same span, we obtain identical $\bar{\beta}$ values. For example, if the maximum range of weights a patient can lift is 200 to 800 grams, the naive estimator will produce identical results based on a design with two weights, 200 grams and 800 grams, and based on three weights, 200 grams, 500 grams and 800 grams.

Following \cite{Lu2015}, the scaling factor for the healthy population can be estimated using a linear mixed effect model based on the 10 subjects in the training data set, with an estimated benchmark value $\beta^{pop}=1.4$N/ms per kilogram of weight increase. Based on these results, we can prescribe a 95\% one-sided confidence interval based on the standard error of this estimator: $[1.27,\infty)$ for the hypothesis testing problem of interest. A naive approach is to compare the scaling factor for the test subject $\bar{\beta}_i$ with the estimated benchmark value along with this confidence interval. If $\bar{\beta}_i$ falls outside the prescribed the confidence interval, the clinicians can be instructed to reject the alternative hypothesis. Albeit simple, this approach is not ideal as the prescribed confidence interval is for the mean scaling factor across a healthy population, and it does not take into account the impact of between subject variability for the purpose of comparison; moreover the within subject (between trial) variability is also not incorporated.
%MAS elevated a footnote:
Such comparison intervals can be improved by instead providing prediction intervals, but these would still be based on the {\it wrong} experimental design (the training subjects' design.)

%\begin{itemize}
%\item $H_0:$ the scaling factor in the test set is the same as in the healthy population represented by the training set: $\delta=0$.
%\item $H_a:$ the scaling factor in the test set is smaller than in the training set. $\delta < 0$.
%\end{itemize}

%Since $\beta^{pop}$ is unknown, the physicians can use the benchmark value [Ying: isn't this 1.4N/ms per 1 kg?] (1.14 N/s per one kilogram weight change) estimated by Lu et al. (need citation) and make a visual judgement on whether the difference $\bar{\delta}=\bar{\beta}-1.14$ is large enough to warrant a warning.

%A slightly more sophisticated method takes into account between trial variability by assigning a rough 95\% confidence interval on $\bar{\beta}$ by prescribing a plus-and-minus two  standard error of $\bar{\beta}$ which can be calculated based on sample variance.
%\begin{eqnarray*}
%\mbox{V}(\bar{\beta})= ...
%\end{eqnarray*}
%One would then compare this range with the 95\% confidence interval for $\hat{\beta}^{pop}$ given in Lu et al. If the first confidence interval falls below the second interval, then the physician can make a judgement that the scaling factor of the test subject is lower than that of healthy population. However this approach is still problematic since: 1) the confidence interval for $\hat{\beta}^{pop}$ is for the population scaling factor, rather than a predictive interval for individual who belongs to the healthy population; 2) We have very little understanding about the behavior of such decisions in terms of false positive and false negative rates.

\subsection{Hierarchical Linear Model via Maximum Likelihood Estimation}\label{HLM}
\setcounter{equation}{3}
An applied statistician might take a modelling approach to compare the scaling factor between the training subjects and the test subject. Since each subject is asked to lift the object over multiple trials and across different weights, we expect that the outcomes will be clustered/correlated within the same subject and the same weight.  As Lu et al.\ (\cite{Lu2015}) pointed out, a linear hierarchical model (\cite{LairdWare1982}) can be used to model this type of data.

%The above hypothesis can be directly tested by building a joint model that combines the training set data and test set data and allows the scaling factor $\beta$ to differ by the training set and every subject in the test set.

Since the training data and the test subject use different experimental designs and a different scaling factor $\beta$ is expected, we outline the model for each group separately using a modified notation.  The following two-level linear hierarchical models, substituting $Y$ for $log(\mbox{PFRL})$ and $W$ for $\mbox{WEIGHT}$, characterize how the observations are generated:
\begin{eqnarray*} \label{eq:two_popns}
\left \{ \begin{array}{l}
Y_{ijt} = \alpha_i + \beta^{pop} W_{ij} + u_{ij}+\epsilon_{ijt}, \, \quad \quad \quad \quad \quad \quad \mbox{(3.1) model for training data} \\
Y_{i'j't'} = \alpha_{i'} + \beta^{test} W_{i'j'} + u_{i'j'}+ \epsilon_{i'j't'},\quad \quad \quad \mbox{(3.2) model for test subject}
\end{array}
\right .  \\
\end{eqnarray*}
for subjects $i$ in the normal or training data, $i=1, \ldots, N^{\mbox{train}}$, and subjects $i'$ in the test data where $i'=1$ in the single-subject design. Subscripts $t$ and $t'$ indicate the trial number for each group, $t=1, \ldots, T$ and $t'=1, \ldots, T'$. $W_{ij}$ and $W_{i'j'}$ specify the weight of the objects being lifted. The terms $u_{ij}$ and $u_{i'j'}$ specify the subject-weight specific effects and they are assumed to follow $N(0, \sigma^2_u)$. The error terms $\epsilon_{ijt}$ and $\epsilon_{i'j't}$ are assumed to be independently distributed $N(0, \sigma^2_\epsilon)$. Note that we are assuming common variances for many of the model components.  We do this to borrow strength from the information gained from the training subjects, but to the extent that test-subject-specific variances wish to be and can be identified, these assumptions can be relaxed.
%aren't we customizing using 'w' design?
%Note that this model specification must be further customized to reflect different experimental designs for data collected for the normal or test populations.
%In particular, repeated measures on the same subject and multiple trials for the same condition must be incorporated into the models, when appropriate.

Instead of testing hypothesis $H_0:\ \beta^{pop}=\beta^{test}$, we consider an equivalent hypothesis $H_0: \delta=\beta^{pop}-\beta^{test}=0$. The quantity $\delta$ can be estimated under the linear hierarchical framework by combining the training data and the test subject, specifying an indicator variable, $\mbox{NEWSUBJ}_i$, set to 1 if subject $i$ is in the test or new population and set to 0 otherwise.  Then, a joint model such as this is fit:
\begin{equation} \label{eq:interaction_model}
Y_{ijt} = \alpha_i + \beta^{pop} W_{ijt} +
\delta W_{ijt}\times\mbox{NEWSUBJ}_i +
\epsilon_{ijt}
\end{equation}

%While the above model represents a simplification of the actual designs used in our study,
The above model can be fit using standard statistical software packages such as \verb SPSS, \verb SAS, \verb Stata  or the \verb nlme  package in \verb R \ (among others) using the Maximum Likelihood Estimation(MLE) approach. Fitting this model will produce point estimates of $\beta^{pop}$ and $\delta$ as well as their standard errors. One can construct a Wald-type statistic (point estimator/standard error) to test whether the difference in scaling factor $\delta$ is significantly different from 0 by comparing it with a $t$-distribution. However, since the Wald test is a large sample based result, when there is only one test subject, it is not appropriate to use the $t$-distribution as the null distribution for the test. As an alternative, permutation based tests are often used for finite samples, but in this example, there are a total of 11 subjects, so the permutation based p-value cannot be smaller than $0.09=1/11$, which implies that the power of the test will be zero if one tries to control the type I error to below 9\%.

Other issues with the MLE approach are that the commonly used software packages tend to be based on strict assumptions about the error structure, such as the same within-subject variation for the training set (healthy subjects) and the test subject (typically patients).

\subsection{Bayesian Hierarchical Linear Model}\label{BayesHLM}

An alternative to the MLE approach is a Bayesian hierarchical model fitting approach \cite{Gelman2013a}. To fit a Bayesian model, one needs to first specify a set of prior distributions (and sometimes hyper-prior distributions) for the parameters of interest. The choice of prior distributions is important as it can have substantial impact on the model. Since the specification and estimation of Bayesian models require a certain level of statistical knowledge, it is less available to practitioners.

To fit (\ref{eq:two_popns}) under the Bayesian paradigm, one could specify the following prior and hyper-prior distributions for the training data,
\begin{eqnarray*}
p(\beta^{pop},\sigma_{\epsilon_1}^2) &\sim & \frac{1}{\sigma_{\epsilon_1}^2} \\
p(\alpha_i) &\sim & N(0, \sigma_{\alpha_1}^2) \\
\sigma_{\alpha_1}^2 &\sim & \mbox{Inv-Gamma}(\eta_1, \nu_1)
\end{eqnarray*}
and for the test subject
\begin{eqnarray*}
p(\beta^{test},\sigma_{\epsilon_2}^2) &\sim & \frac{1}{\sigma_{\epsilon_2}^2} \\
p(\alpha_{i'}) &\sim & N(0, \sigma_{\alpha_2}^2) \\
\sigma_{\alpha_2}^2 &\sim & \mbox{Inv-Gamma}(\eta_2, \nu_2)
\end{eqnarray*}

Notice here that the Bayesian model allows us to explicitly specify the variance components differently for the training data and the test subject. On the other hand, this setup also allows one to borrow strength and estimate the variance components jointly by forcing $\sigma_{\alpha_1}^2=\sigma_{\alpha_2}^2$ and $\sigma_{\epsilon_1}^2=\sigma_{\epsilon_2}^2$. Without loss of generality, we do this in most of what follows.

In the current setup, we choose to apply non-informative priors to avoid the impact of prior influence on model estimation \cite{Gelman2013a}\cite{stan-manual2015}, with the exception of the subject-specific intercept variance. Since there are only 10 training subjects and one test subject, we decide to apply an informative prior, the inverse-gamma distribution, on the variance of $\alpha_i$.
This prior ($\mbox{Inv-Gamma}(\eta, \nu)$) is determined by two parameters $\eta$ and $\nu$, corresponding to a prior mean variance value $\nu/(\eta-1)$.  Our choices for $\nu, \eta$,  correspond closely to the MLE estimate of the variance.

The Bayesian approach then estimates the parameters via posterior simulation. Let $\Theta=(\beta^{pop}, \beta^{test}, \sigma_{\alpha_1}^2, \sigma_{\alpha_2}^2, \sigma_{\epsilon_1}^2,\sigma_{\epsilon_2}^2)$, then the posterior distribution can be derived using Bayes formula,
\begin{eqnarray*}
p(\Theta | Y^{train}, Y^{test},W^{train},W^{test})
 &=& \frac{p(Y^{train}, Y^{test}|\Theta,W^*)p_0(\Theta)} {\int p(Y^{train}, Y^{test}|\Theta,W^*)p_0(\Theta) d\Theta} \\
&\propto & p(Y^{train}, Y^{test}|\Theta,W^*)p_0(\Theta)
\end{eqnarray*}
where $W^*=\{W^{train},W^{test}\}$ and $p_0(\Theta)$ is the prior distribution of the parameters. When the closed-form of the posterior distribution is not available, it can be approximated using Monte-Carlo simulations and used for inference \cite{rstan-software2015}\cite{gelman2007data}.

For this example, with a simple reparameterization, $\beta^{test}=\beta^{pop}-\delta$, we can generate the posterior distribution of $p(\delta|Y^{train}, Y^{test})$. Unlike the MLE approach, in which a single point estimator $\hat{\delta}$ is produced for the quantity of interest, the posterior distribution of $\delta$ is the basis of Bayesian inference. For example the maximum a posteriori probability (MAP) value is the $\delta$ value corresponding to the peak of the posterior distribution and can be viewed as a Bayesian version of the point estimator, and the posterior standard deviation is a measure of the variability in $\delta$ (there is no sampling distribution of an estimator; instead, there is a posterior for the corresponding parameter).

Bayesian hypothesis testing does not come naturally due to the fundamental difference in problem formulation--the Bayesian approach posits that all parameters are random variables and follow a distribution, which is estimated by the posterior distribution, while the Neyman-Pearson type of hypotheses typically focus on a single parameter value.  Meng (\cite{Meng1994}) proposed the posterior $p$-value, which is the Bayesian counterpart of the classical $p$-value, by simulating the joint posterior distribution of replicate data and the (nuisance) parameters, both conditional on the null hypothesis and calculating the tail-area probability of a ``test statistic" using this distribution. However, the posterior $p$-value has been criticized for its tendency to center around 0.5 for the hierarchical model \cite{Gelman2013a, Gelman2013b}. Instead, as shown in Section~\ref{sec2}, we consider a shorthand way of evaluating the posterior probability of $\delta<0$. Namely, for a test at level 5\%, if $p(\delta<0 | Y^{train}, Y^{test})<0.05$, then we reject the null hypothesis.  In other words, in this instance, there is sufficient evidence (95\% of the posterior mass) to support $\delta>0$, or deviation from the training data.

\section{A New Paradigm For Single-subject Design Analysis}\label{sec2}
Unlike directly comparing the naive estimator in equation (\ref{eq:nparm0}) with a predetermined benchmark value to make a visual judgement about the status of the test subject, the use of Maximum Likelihood and Bayesian modeling allow us to compare the test subject with the training data set while taking into account the within-subject and between-subject variability, and statistical tests are available to assist decisions under these modelling frameworks. However these approaches are not practical in the clinical setting. In order to make an inference regarding a new subject, one needs to refit the entire multilevel model, which is time consuming and not user-friendly in the clinical setting. Without proper training in statistics, these methods are practically unavailable to clinicians. Moreover since most of the statistical parametric modelling approach depends on a large sample, the behavior of the aforementioned methods in hypothesis testing regarding a single subject is unknown.\footnote{The Bayesian approach handles the non-asymptotic setting more elegantly, but is inherently more difficult to fit without specialized knowledge of statistical programming languages such as STAN, BUGS or JAGS.} The error rates such as false positive and false negative rates associated with the decisions can be off target.

%The posterior distribution of $\delta$ suffers from the sampling variability that is inherent to the single subject, which is likely to be very large (making point estimates imprecise); moreover,
%the
%in the Bayesian setting, we're not talking about asymptotics.
%distributions
%asymptotics
%upon which most tests of significance are based may not hold for the small sample case of N of 1 subject.  False positive and negative rates may be other than the target.
%\end{enumerate}
%this is the old first paragraph.  Needs updating.  I give an attempt following it.
%Up to this point, we have presented a mathematical version of the physiological relationships, but this was somewhat simplified.  For the population of normal subjects discussed here, we have repeated trials for varying conditions (weights) in order to reduce measurement error, and statistical models must extend the model above to include correlation within condition and subject. We will formalize this approach, which requires fitting a rather sophisticated linear hierarchical model, in the next section.  We note, however, that fitting these more complex models is not a practical tool for clinicians, and our inferential concerns remain in the multilevel setting.

%\subsection{Alternative Approach}

To address these concerns, we propose a novel approach that will allow clinicians without any formal statistical training to make an informed decision about the test subject's status as compared to reference subjects in the training data. We first outline the general ideas, and then present the detailed algorithm for the method.

We start with the naive estimator $\bar{\beta}^{test}$ based on (\ref{eq:nparm0}), which is available in the clinical setting. The goal is to provide the clinician with a template distribution of the possible values that we expect to observe given the weight design and the number of repeated measures used by the test subject. This template distribution is to be developed in a laboratory setting where the scientists and statisticians collaborate to design experiments and collect data based on a carefully controlled set of training subjects, for example, a random sample of healthy subjects. Based on the template distribution, the clinician can easily test the hypotheses such as
\begin{description}
\item[H0: Patient has healthy anticipatory control]$\beta^{test}=\beta^{pop}$.
\item[Ha: Patient does not have healthy anticipatory control]   $\beta^{test}< \beta^{pop}$
\end{description}

The probability of observing any values $\beta^{test}$ as extreme as the naive estimate $\bar{\beta}^{test}$ had the test subject behaved the same way as the reference population
%represented by the training subjects
can be easily generated using the template distribution. This probability has the interpretation of a classic $p$-value in a statistical inference problem (P(observation as extreme, given the model) under the null). The clinician can choose a desired level of the test, say $0.05$, and reject the null hypothesis whenever the $p$-value is less than $0.05$.
%YING: I took out alpha since we use alpha otherwise elsewhere.
An equivalent alternative is to compare the naive estimate $\bar{\beta}^{test}$ directly with the critical value $C_{0.05}(\beta^{test})$ derived from the template distribution. If $\bar{\beta}^{test}<C_{0.05}(\beta^{test})$ then reject the null hypothesis.
Moreover, since the template distribution is derived in a laboratory setup, the performance of such decisions can be evaluated ahead of time. Along with a convenient test, the expected error rates (or equivalently, the power) associated with the decision will also be reported.

\subsection{The Algorithm for Deriving the Template Distribution} \label{algorithm}

To derive this template distribution, we exploit a feature of Bayesian modeling, which is that posterior predictions are easily computed using any combination of model parameters.  Crucially, this allows one to vary the design between training and test subjects and propagates parameter uncertainty into the predictions, providing a natural framework for statistical inference that does not rely on asymptotic theory.  The details are given below:

\begin{enumerate}
\item We fit a Bayesian hierarchical model (3.1) using the training data set alone to obtain the posterior distribution of the parameters.
%from a Bayesian model fit to the training dataset.
For model~(3.1), these parameters are
$\Theta=\{a,\beta^{pop},\sigma^2_{\alpha},\sigma_{u}^2,\sigma^2_\epsilon\}$, for which we label the posterior $h(\Theta)=p(\Theta|Y^{train})$.
\item Given a new design $W^{new}$,
%the number of trials used in model~(3.2), say,
we assume, under the null, that all parameters $\Theta$ in model~(3.2) are the same.  We then
generate a set of posterior predictive outcomes $\tilde{y}\sim \mbox{MVN}(\mu^{new},\Omega^{new})$,
where for this model $\mu^{new}=a+\beta^{pop} W^{new}$ and $\Omega^{new}$ has compound symmetry structure induced by the random effects for intercept and trials.
Specifically,
\begin{equation*}
\{\Omega^{new}\}_{ijt,i'j't'}=
  \begin{cases}
       0   & \quad \text{if } i\neq i' \\
       \sigma^2_\alpha  & \quad \text{if } i=i',j\neq j' \\
       \sigma^2_\alpha + \sigma^2_u  & \quad \text{if } i=i',j=j',t\neq t' \\
        \sigma^2_\alpha + \sigma^2_u + \sigma^2_\epsilon & \quad \text{if } i=i',j=j',t=t'
  \end{cases}
\end{equation*}
This posterior, $p(\tilde{y}|\Theta, W^{new})$ can be viewed as the distribution of the future outcomes that would be observed, were the new design applied to subjects from the training sample.\footnote{Said differently, these predictive outcomes will inherit the hierarchical structure (subject/weight/trial) based on the new design, while also inheriting the behavior of the reference subjects in the training set.}
By generating these pseudo-outcomes in a Bayesian framework, the model uncertainty is propagated from $h(\Theta)$ to the predictions, $\tilde{y}$, representing our current understanding of the physiological process, and this can be updated should more training subjects become available.
\item In order to construct the template (reference) distribution, we need a large sample of pseudo-subjects drawn from $p(\tilde{y}|\Theta, W^{new})$. For any new subject $i$, we take $\tilde{y}_i$ and compute $\tilde{\bar{\beta}}$ using equation (\ref{eq:nparm0}).  This is the natural single-number (non-parametric) summary statistic one would compute for a new subject. The density of $\tilde{\bar{\beta}}$ over a large number of such pseudo-subjects is an approximation to the distribution for a randomly drawn new subject under the null hypothesis. Given a predetermined false positive rate (FPR), say 5\%, the critical value for rejecting the null hypothesis can be obtained empirically.
\item Recall that $\beta^{test}=\beta^{pop}+\delta$.  As in step 1, the posterior predictive draws of $\tilde{y}$ under a specific alternative value of $\delta$, call it $\delta_{alt}$, can be obtained by a location shift of the posterior predictive distribution of $\tilde{y}$ by $\delta_{alt}$, again, due to
the linearity and normality of the models examined. Then, following steps 2 and 3, using the shifted $\tilde{y}_{alt}+\delta_{alt}$, we can easily obtain distributions of  $\tilde{\bar{\beta}}_{alt}$ under different hypothesized alternative values of $\delta_{alt}$. These distributions can be used to assess the false negative rate (FNR) or power of the decisions made regarding to the hypothesis.
%
%      and a collection of posterior draws $\tilde{y}_j$, we can approximate the posterior distribution of $\delta$ for a give value $\tilde{y}_j$ as $\frac{g(\tilde{y}_j,\delta_k)}{\sum_k g(\tilde{y}_j,\delta_k)}$.  Note that $\phi$ is the normal density, $N_1$ are the posterior draws of the parameters and $N_2$ are the posterior predictive draws of $y$. This is a slight simplification, in which $\sigma_i^2$ includes all error variances $\sigma^2_\alpha, \sigma_u^2,\sigma^2_\epsilon$, and it assumes independence across observations within the same subject, when in fact there is dependence induced by the $\alpha_i$ and $u_{it}$ terms across the N(conditions) $\times$ N(trials).
%
\end{enumerate}

\section{Simulation Studies}\label{sim}
In this section, we conduct a set of simulations to assess the performance of the proposed method.  We apply the method to the hypothesis testing problem of whether a (new) test subject has healthy predictive control of fingertip forces to object weight during precision grasp. This test is based on the naive estimator $\bar{\beta}^{test}$ and our Bayesian model-based template distribution. For comparison, we will also assess the performance of a Bayesian hierarchical model and the Wald test for the MLE estimator $\hat{\delta}$ in a linear mixed effects model.

\subsection{Setup}\label{setup}
First we outline the basic simulation setup. Multiple samples of training data and test subjects will be simulated according to the following data generating process:
\begin{description}
\item[Training data] The training data contains 10 subjects simulated from model (3.1), with the scaling parameter $\beta^{pop}=1.4$. The other parameter values are set as follows, $a=2.8$, and $\sigma_{\alpha_1}=0.3$, $\sigma_{u_1}=0.1$ and $\sigma_{\epsilon}=0.2$. This set of parameters are assumed to be the population parameters of the healthy subjects.

The design matrix for the training data assumes that each subject lifts 10 weights, ranging from 250 grams to 750 grams, 50 grams apart. The subjects lift each weight over 6 trials (after an initial practice trial that is discarded).
\item[Test subject] The test subject is simulated following model (3.2) with the same parameter values as the training data set except for $\beta^{test}$.
A range of scaling factors $\beta^{test}$ from $1.4$ to $0.1$ are used, covering both the null and alternative parameter space. The design matrix used for the test subject is different from that used in the training set. Since the test subjects will typically be patients, a less involved weight design and fewer repeated trials will be used. For test subjects, the number of trials after practice is set to 5.
We examine two weight design scenarios 1 and 2, defined respectively as:
\begin{enumerate}
\item Each subject lifts only two weights at 250 grams and 500 grams
\item Each subject lifts three weights, at 250, 500 and 750 grams
\end{enumerate}

%\begin{itemize}
%\item When $\delta=0$, the test subject comes from the same distribution as the subjects in the training set.  This is H0.
%\item When $\delta <0$, the test subject does not scale as much as the training subjects. These are values for Ha.
%Notice here we assume a big deviation, as clinically it is more meaningful the lack of scaling (i.e. do not scale, rather than less scale).
%    \begin{itemize}
%    \item There is a biomechanical interpretation: once the subjects developed feed forward control, given a narrow range of lifting time, the rate of load force must be proportional to the weight. In contrast, when there is lack of feed-forward control, we tend to see there is lack of proportional change.
%    \end{itemize}
%\end{itemize}
%\end{itemize}
\end{description}
%\begin{adjustwidth}{2cm}{}
%\end{adjustwidth}

When the test subject is simulated under $\beta^{test}=\beta^{pop}=1.4$, it corresponds to the null hypothesis. When the test subject is simulated under $\beta^{test}<1.4$, it corresponds to a case within the alternative hypothesis parameter space. The purpose of the simulation studies is to compare the performance of different methods in terms of false positive rate and false negative rate. To estimate the false positive rate and false negative rate, we generate a large number\footnote{We typically generate 1000 replicates.} of replicates of the training and the test sets as necessary.

For each replicate of a simulated data set, we consider four different ways of estimating the parameters of interest and testing the null and alternative hypotheses.
\begin{description}
%YING: I reordered these
\item[Test A: naive estimator $\bar{\beta}$ and its sampling distribution] For a single simulated test subject, we calculate $\bar{\beta}^{test}$ using (\ref{eq:nparm0}). We denote this estimator $\bar{\beta}^{test}_0$ when the test subject is simulated under the null hypothesis (when $\beta^{test}=1.4$). The distribution of 1000 samples of $\bar{\beta}^{test}_0$s approximates the sampling distribution of $\bar{\beta}^{test}$ under the null hypothesis, denoted by $Dist(A)$.

A decision rule is proposed using $Dist(A)$: for a one-sided test at level $\alpha$, the rejection region is $\mathcal{D_{\alpha, A}}=\{\bar{\beta}: \bar{\beta}<C_{\alpha, Dist(A)} \}$, where $C_{\alpha, Dist(A)}$ is the $\alpha$th percentile value of $Dist(A)$. By directly comparing a new subject's naive estimator $\bar{\beta}$ with this rejection region, a decision can be made regarding a single test subject.

When the hypothetical test subject is simulated using a range of values $\beta^{test}=\beta^{pop}-\delta$, where $\delta=(0.1, 0.2, \ldots, 1.3)$, it forms the alternative hypothesis space. For each value of $\beta^{test}<1.4$ so derived, we can estimate the false negative rate by summarizing test results across 1000 copies of new test subjects. Note that the false positive rate for this test is $\alpha$ by construction.

To construct this test, we have a model for the data generating process, which we assume is a close approximation to what we would observe in a large population of normals.
Since the decision rule is based on the sampling distribution of the test statistic $\bar{\beta}^{test}$ under the null, and the nature of this test is non-parametric, we will use the error rates derived based on this test as the ``gold standard'' for comparison purposes.

\item[Test B: the estimator $\bar{\delta}$ and Bayesian-based template distributions]
Following the method outlined in section \ref{algorithm}, Using the R Bayesian package \texttt{rstan}\cite{rstan-software2015}, we first fit model~(3.1) to training data with non-informative priors on the hyperparameters, except for $\sigma_\alpha^2$. The prior values for the hyperparameters are set to $\eta=5, \nu=1$.  Three Markov Chain Monte Carlo (MCMC) chains were run, each of 2000 draws.
%NOTE: I used fewer draws out of convenience....  FIX??? as long as converges, shorter chain is ok with me
The first 1000 draws of each chain are burn-in and are discarded.\footnote{Convergence of the chains was evaluated using diagnostic statistic $R$ as described in \cite{gelman1992inference}.
All runs converged with $R$ extremely close to 1.}
Based on the Bayesian fit, we then generate 3000 posterior predictive draws of $\tilde{y}$ using the test subject design.
It is important to note here that we are ``borrowing'' the design used in a future observation as a template, but we do not actually use any real future observations in the construction of the reference distribution (as opposed to the Bayesian hierarchical modeling method, which does estimate a $\delta$ from the test data).
The template distributions are derived following steps 1-4 in the proposed algorithm in section \ref{algorithm}. In particular, we denote the template distribution under the null hypothesis $Dist(B)$.

Note that each set of training data generates a complete template null.  We repeat this generative process across different training data to understand the variability inherent in the Bayesian analysis.  In practice, a single template null will be used.

A decision rule is proposed using $Dist(B)$: for a one-sided test at level $\alpha$, the rejection region is $\mathcal{D_{\alpha, B}}=\{\bar{\beta}: \bar{\beta}<C_{\alpha, Dist(B)} \}$, where $C_{\alpha, Dist(B)}$ is the $\alpha$th percentile value of $Dist(B)$. By directly by comparing the naive estimator $\bar{\beta}$ with this rejection region, a decision can be made regarding a single test subject. Similarly, the error rates associated with this test can be summarized when using different true $\beta^{test}$ values for the alternative.\footnote{For the evaluation of Type I and II error, we average the error rates associated with 1000 unique training sets and their corresponding template distributions. In this evaluation, test subjects are generated from the known distribution described in A. For this simulation setup, we can also evaluate the variability of these error rates.}

This is also the assessment we propose for the clinical set-up. In the clinical setup where the empirical error rates are not available, but can be calculated based on examining the overlapping areas between the template distributions under the null and under a specific alternative parameter value. In the simulation studies, we will evaluate this more directly applicable variant on $Dist(B)$, calling it $Dist(B*)$.

%be  using model (\ref{m21}) and the design matrix of model (\ref{m22}). Viewing this set of posterior draws as realizations of the outcomes from the population represented by the training set (and the inherent variability of parameters induced by the study design), we then compute the posterior density of N-of-1 subject values $\delta$ based on steps 1-4 (the MAP of individual-level posteriors) as described in the previous section. $Dist(B)$.
\item[Test C: Bayesian posterior $p$-value] For each pair of training set and single test subject, we fit a joint Bayesian hierarchical model~(\ref{eq:two_popns}) as in section~\ref{BayesHLM}. Based on the posterior distribution of $\delta$, we can compute the probability $p(\delta\leq 0|Y^{train}, Y^{test})=p(\beta^{test}\geq\beta^{pop}|Y^{train}, Y^{test})$. This probability can be interpreted as a $p$-value,
under the Bayesian framework as the ``support'' of the hypothesis $\delta=0$ under a one-sided alternative (when the support drops below 0.05, say, we reject the supposition that $\delta$ is not positive; as a reminder, when $\delta>0$, $\beta^{test}<\beta^{pop}$).  %A posterior $p$-value summarizing the support for $\bar{\beta}<\beta^*$, for some meaningful cut point $\beta^*$ can be calculated based the posterior predictive draws (\cite{Meng1994}) utilizing the Bayesian fitting of model~\ref{eq:two_popns}, previously described.
Across 1000 copies of simulated datasets for each alternative (paired with test sets), we can approximate the false negative rate of the posterior $p$-value based test and average these.
\item[Test D: Wald test for MLE estimator $\hat{\delta}$] For each pair of training set and single test subject, we estimate the difference in scaling factor between the test subject and the population (as represented by the training data) using model (\ref{eq:interaction_model}), which yields the difference estimate $\hat{\delta}$. We will use the \texttt{R}~package \texttt{nlme}~to make this estimate. The \texttt{nlme}~package also reports a Wald test statistic $\hat{\delta}/\hat{SE}(\delta)$, and a $p$-value is calculated based on this test statistic assuming it has a $t$ distribution under the null with a degree freedom based on the hierarchical linear model framework (\cite{LairdWare1982} and \cite{PinheiroBates2000}). The result of this test at level $\alpha$ is also summarized across 1000 pairs of simulated training set plus a single test subject to approximate the false positive and false negative rates.
\end{description}

\subsection{Results}\label{sim_results}
In this section, we summarize the results of the simulations.  Our goals include understanding the performance of the proposed test by comparing a naive estimator $\bar{\beta}^{test}$ with the Bayesian-based template distribution (Test B). The false positive rate and false negative rate of this proposed test will be compared with those based on Test A (``gold standard''). Since in practice, the distribution of test subjects under the alternative is not known, the use of ``gold standard'' test subjects is an idealized evaluation, which we report for Test B.
%I don't think bold font works in final draft.
With real data (rather than multiple copies of simulated data), the error rates for Test B are not directly obtainable, but the expected false negative rate can be approximated using a direct method based on calculating the overlapping area between the template distributions under the null and then an alternative constructed via a location shift of the null, and it is reported under column ``Test $B*$''.
The model-based results Test C and Test D will also be assessed to understand the behavior of these tests in small samples.
%These results are summarized in Table 1.

\begin{table}\label{err_table2}
%\begin{adjustwidth}{-2.25in}{0in}
%\scriptsize{
\small{
\begin{tabular}{c|ccccc|ccccc}
\hline
 & \multicolumn{5}{c}{Scenario One} & \multicolumn{5}{c}{Scenario Two} \\
$\delta_{alt}$ & Test A & Test B & Test $B*$ & Test C & Test D & Test A & Test B & Test $B*$ & Test C & Test D\\
\hline
\\
& \multicolumn{10}{c}{Level of Test: 5\%} \\
   0.0 & 0.050 & 0.056 & 0.050 & 0.043 & 0.250 & 0.050 & 0.043 & 0.050 & 0.039 & 0.326 \\
   0.1 & 0.934 & 0.936 & 0.935 & 0.935 & 0.660 & 0.895 & 0.922 & 0.917 & 0.920 & 0.623 \\
   0.2 & 0.915 & 0.928 & 0.916 & 0.915 & 0.609 & 0.835 & 0.868 & 0.870 & 0.868 & 0.552 \\
   0.3 & 0.893 & 0.882 & 0.895 & 0.885 & 0.597 & 0.777 & 0.785 & 0.807 & 0.791 & 0.467 \\
   0.4 & 0.859 & 0.871 & 0.869 & 0.858 & 0.535 & 0.673 & 0.722 & 0.729 & 0.690 & 0.451 \\
   0.5 & 0.828 & 0.841 & 0.839 & 0.818 & 0.486 & 0.592 & 0.633 & 0.637 & 0.599 & 0.360 \\
   0.6 & 0.792 & 0.820 & 0.804 & 0.786 & 0.431 & 0.488 & 0.530 & 0.537 & 0.496 & 0.256 \\
   0.7 & 0.754 & 0.763 & 0.766 & 0.739 & 0.357 & 0.391 & 0.421 & 0.435 & 0.369 & 0.224 \\
   0.8 & 0.715 & 0.732 & 0.724 & 0.673 & 0.343 & 0.287 & 0.338 & 0.337 & 0.269 & 0.179 \\
   0.9 & 0.668 & 0.666 & 0.678 & 0.611 & 0.316 & 0.202 & 0.246 & 0.250 & 0.195 & 0.136 \\
   1.0 & 0.618 & 0.633 & 0.630 & 0.546 & 0.251 & 0.133 & 0.160 & 0.177 & 0.116 & 0.086 \\
   1.1 & 0.581 & 0.592 & 0.579 & 0.486 & 0.219 & 0.080 & 0.112 & 0.119 & 0.070 & 0.072 \\
   1.2 & 0.516 & 0.546 & 0.527 & 0.430 & 0.193 & 0.052 & 0.069 & 0.077 & 0.042 & 0.044 \\
   1.3 & 0.461 & 0.455 & 0.475 & 0.363 & 0.156 & 0.029 & 0.044 & 0.047 & 0.021 & 0.042 \\
\hline
\\
& \multicolumn{10}{c}{Level of Test: 10\%} \\
   0.0 & 0.100 & 0.103 & 0.100 & 0.093 & 0.307 & 0.100 & 0.086 & 0.100 & 0.089 & 0.366 \\
   0.1 & 0.879 & 0.868 & 0.875 & 0.877 & 0.614 & 0.824 & 0.848 & 0.846 & 0.847 & 0.566 \\
   0.2 & 0.852 & 0.851 & 0.846 & 0.851 & 0.562 & 0.739 & 0.772 & 0.777 & 0.760 & 0.513 \\
   0.3 & 0.818 & 0.797 & 0.812 & 0.814 & 0.527 & 0.669 & 0.663 & 0.692 & 0.668 & 0.437 \\
   0.4 & 0.768 & 0.769 & 0.775 & 0.772 & 0.478 & 0.551 & 0.595 & 0.596 & 0.571 & 0.403 \\
   0.5 & 0.726 & 0.739 & 0.734 & 0.727 & 0.441 & 0.467 & 0.496 & 0.494 & 0.456 & 0.314 \\
   0.6 & 0.681 & 0.696 & 0.689 & 0.661 & 0.379 & 0.355 & 0.390 & 0.393 & 0.336 & 0.218 \\
   0.7 & 0.629 & 0.638 & 0.641 & 0.593 & 0.304 & 0.273 & 0.297 & 0.298 & 0.242 & 0.184 \\
   0.8 & 0.595 & 0.600 & 0.590 & 0.529 & 0.296 & 0.183 & 0.211 & 0.216 & 0.168 & 0.147 \\
   0.9 & 0.539 & 0.528 & 0.539 & 0.469 & 0.273 & 0.124 & 0.141 & 0.150 & 0.104 & 0.111 \\
   1.0 & 0.485 & 0.481 & 0.486 & 0.402 & 0.218 & 0.079 & 0.082 & 0.099 & 0.060 & 0.073 \\
   1.1 & 0.445 & 0.450 & 0.434 & 0.334 & 0.179 & 0.042 & 0.060 & 0.062 & 0.033 & 0.058 \\
   1.2 & 0.381 & 0.400 & 0.384 & 0.278 & 0.153 & 0.027 & 0.030 & 0.037 & 0.016 & 0.036 \\
   1.3 & 0.328 & 0.334 & 0.335 & 0.236 & 0.124 & 0.013 & 0.021 & 0.021 & 0.010 & 0.033 \\

\hline
\end{tabular}
}
%\end{adjustwidth}
\caption{The comparison of error rates for scenarios with different weight conditions. The first column shows the true values of $\delta_{alt}$ under which test data were generated. Two desired levels of test are considered, 5\% and 10\%. The first row shows the false positive rate of different tests. The remaining rows show the false negative rate since $\delta_{alt}>0$. Since Test B is to be used in the clinical setting, the expected false negative rate associated with a given level of the test are shown in column ``Test $B*$'' (see text for further detail).}
\end{table}

Table 1 shows the simulated error rate under various hypothesized $\delta_{alt}=\beta^{pop}-\beta^{test}$ values using tests A-D. In the first row when $\delta_{alt}=0$, the test subjects are generated under the null hypothesis space, hence the corresponding quantities are False Positive Rates (FPR) of different tests. They are also the type I error rate. When $\delta_{alt} >0$, the test subjects are generated under the alternative hypothesis space hence the remaining rows report the false negative rates (FNR or 1-power).

Focusing first on Test A for Scenario One when two weights are used, with a level 0.05 test, we see that the FPR is 0.05 by construction (the critical value was based on the 5th percentile on the same distribution).  FNRs are quite high at 46.1\% even for a very strong alternative, as is given by the last entry  ($\delta_{alt}=1.3$).  This is to be expected with a small test sample and a low FPR.  Continuing to observe Test A, we see that under Scenario Two (right column) when three weights across a greater range are used, the False Negative Rate decreases substantially.\footnote{As we have noted, for two or three equispaced conditions, the naive estimator is determined by observations at the two extreme weights (other terms cancel). This is revisited in the power analysis to follow.  For now, we note that these findings for $Dist(A)$ (and $Dist(B))$ are driven by the increased distance between the extreme weight conditions.}
At $\delta=0.7$, when the scaling factor of the test subject is half of that of the expected value of normal training subjects, the FNR is 39.1\%. At greater $\delta$ values,  the extra weight range makes a big difference in terms of reducing the false negative rate and increasing the potential power of this test.
% the strongest alternative now yields a FNR of 2.9\% even for a level 0.05 test.

In comparison, the error rates of our proposed method Test B, using only the training data and the test design, are only slightly higher than those of Test A when the true sampling distribution of the scaling factor $\beta^{test}$ is assumed to be known. In particular, the false positive of Test B is 5.6\% suggesting that our proposed test is capable of controlling the type I error at the desired level.

Since in practice the expected error rates of Test B are unknown, under column Test $B*$, we report error rates approximated by calculating the overlapping area of the null template distribution and the alternative template distribution when $\delta=\delta_{alt}$ (derived using the method in Section~\ref{algorithm} step 4). Both tests B and $B*$ yield strikingly similar FPRs and FNRs under our range of scenarios, and these are also quite close to what we would expect from the standard in Test A.  In addition, we can estimate the standard error of Test B's and $B*$'s FNRs, which range from 0.01 to 0.06 in nearly every instance, with smaller s.e.\ for Scenario Two. The availability of these standard errors allows us to report, in a practical setting, an estimated error rate with a confidence interval for Test B using those quantities calculated under Test $B*$.

With only one test subject, we expect the power of the test will be low.
However, when the level of the test is set to be 10\%, the false negative rate is greatly reduced. When $\delta>0.7$, the False Negative Rate under Scenario Two for Test A is less than 20\%, corresponding to a power of at least 80\%, and the performance of Test B and Test $B*$ continue to be very similar.

%The practical surrogate test $B*$, we note that the former uses samples from the ideal test set, while the latter establishes the proportion of distributional overlap, which corresponds more closely to the clinical setting in which information about the alternative distribution is basically unknown.

The performance of model-based tests C and D are also assessed in Table 1. Surprisingly, we find that Test D, the classical mixed effects model approach, fails to control the type I error rate at the proper level. For example, under scenario one, the false positive rate for Test D is 25\% while the desired rate is 5\%.  This suggests that the Wald test relies strongly on asymptotic behavior and it is not appropriate for a single-case design when the sample size is very small.

Test C, on the other hand, does a better job at controlling the FPR and manages to achieve FNRs that are comparable to the standard set by Test A.  Since the $p$-values of Test C are evaluated empirically based on the posterior distribution of $\delta$, it is free from the small sample ``curse.''
%as in Test D.
Indeed, the fully Bayesian hierarchical model performs a little better than Test A with respect to FNR under scenario one, which is when the information collected from a test subject is more limited. This suggests that, if {\it modelled under the correct data generating process}, by jointly modeling the training data (under a stronger design) and the test subject, a slightly higher powered test can be attained.
%than Test A using only the sampling distribution of the test subjects.

Figure~\ref{null} compares the sampling null distribution of $\bar{\delta}$ ($Dist(A)$) and the proposed alternative null distribution based on Bayesian posterior predictive draws ($Dist(B)$). Since multiple samples of $Dist(B)$ are available, we can display the variability inherent in model estimation by superimposing 100 samples of this distribution upon the standard $Dist(A)$.
The left panel plots the results based on the two-weight design, and the right panel plots the results based on the three-weight (more extreme weight) design. We see that the two distributions are fairly close, and the variability in $Dist(B)$ is surprisingly small, considering that it is estimated from only 10 normal subjects. This is because the training data uses a stronger within-subject design where each subject lifts 10 different weights with more repeats.

\begin{figure}[ht]
\begin{center}
%%\framebox{\includegraphics[width=2in]{normL.png}}
\includegraphics[width=2.5in]{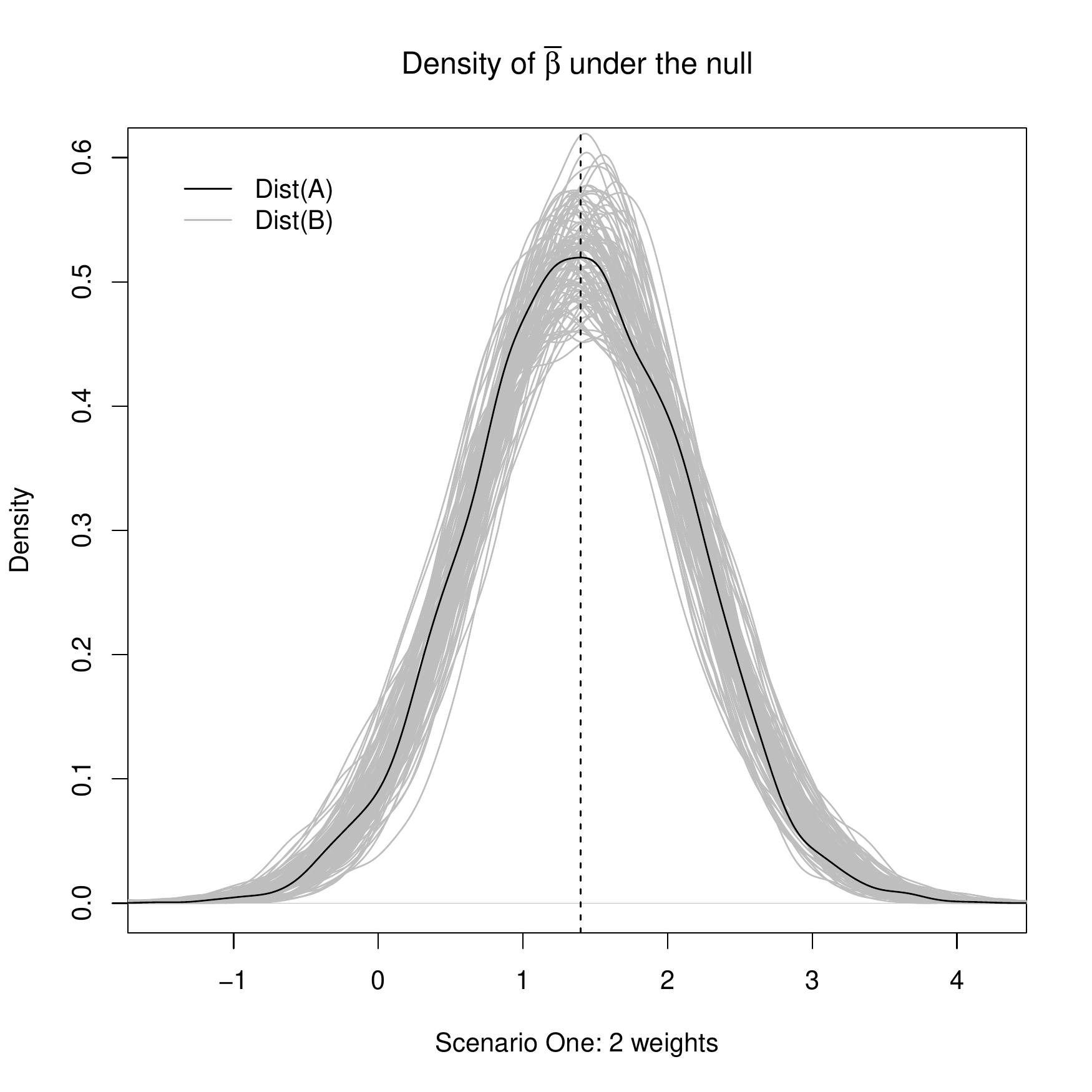}
\includegraphics[width=2.5in]{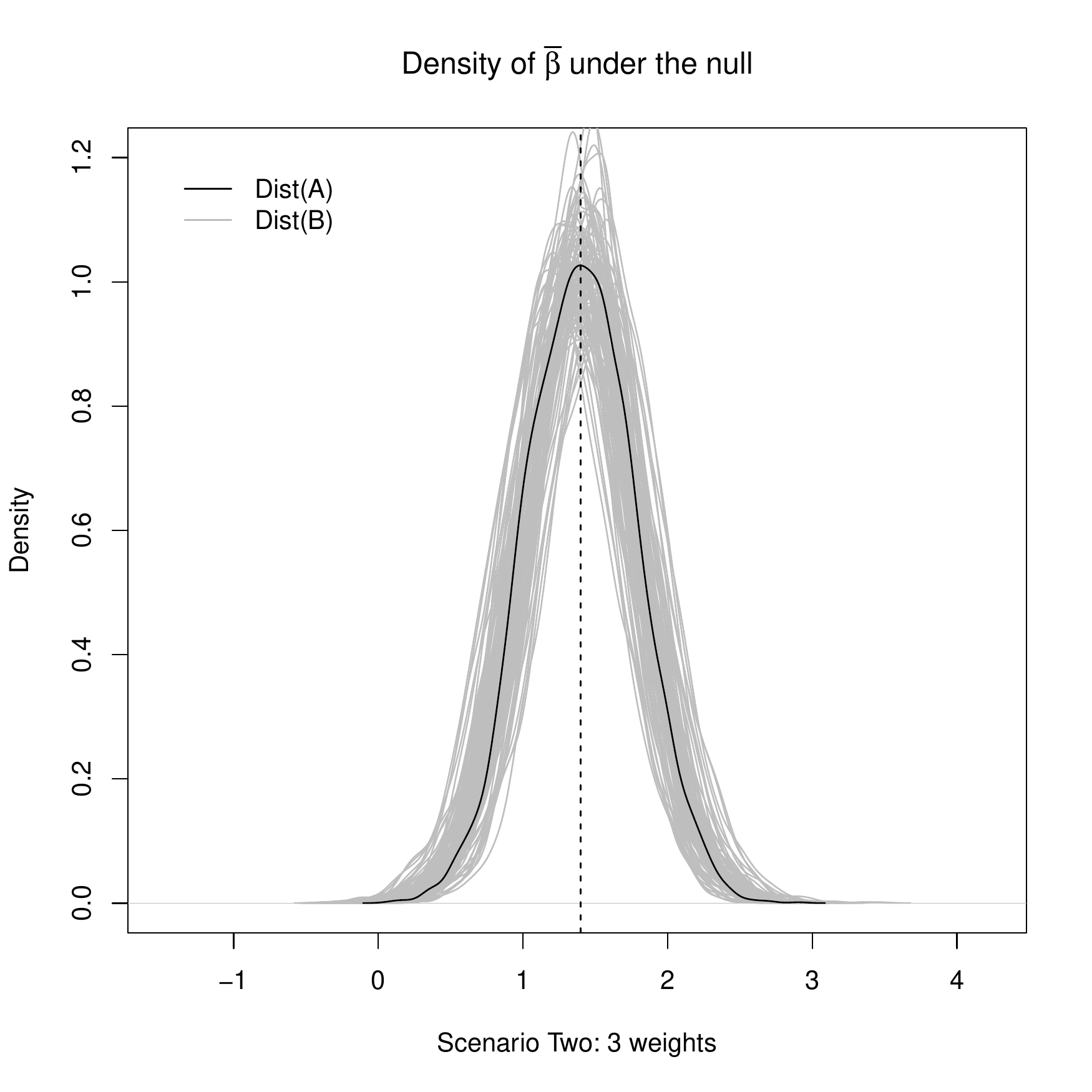}
\caption{$Dist(A)$ is the true density of $\bar{\beta}$ under the null hypothesis. In grey are 100 realizations of $Dist(B)$ under the null based on the proposed method using a single copy of simulated training set as the basis for the realized density estimate.}
\end{center}
\end{figure}\label{null}

\subsection{Power Analysis}

To understand the impact of the experimental design for the single test subject, we examined a range of plausible scenarios.  In each study, we estimate the template null distribution $Dist(B)$ using our method, and the alternative distribution by shifting the location of $Dist(B)$ by $\delta_{alt}$ units. The level of the test is set to be 10\%. The false negative rate is then calculated by the area under the alternative $Dist(B)$ that falls into the rejection region for a test of given level, in this case, the rejection region is
$\mathcal{D_{\alpha, B}}=\{\bar{\beta}: \bar{\beta}<C_{\alpha, Dist(B)} \}$, with $\alpha=0.10$.
%$\{ C_{0.1, Dist(B)}\}$.
For each scenario, the power of the test is calculated as 1-(false negative rate) averaged over 100 different estimation runs.  In one set of simulations, we vary the size of the training set from 10 to 20 subjects, crossed with varying the number of trials from 5 to 10, holding the previously explored weight conditions (250g, 500g, 750g) constant.  The only notable impact was increasing the number of trials to 10 per condition.  What this suggests is that 10 subjects (under the more complex design outlined in our simulation study) capture sufficient information about the population of normals.\footnote{Clearly, the large number of conditions and repeated trials in the training set contributes greatly to the stability of the estimates.} We do not report this set of comparisons, but simply note that the training set sample size is of secondary importance.

In the second study, the number of training subjects was fixed at 10 and the number of test subject trials was fixed at 5.  What vary are the number and range of conditions, which are given by these six scenarios: (200g and 400g);  (200g and 600g);  (200g, 400g, 600g);  (200g and 800g);  (200g, 500g, 800g);  (200g, 400g, 600g, 800g).  Given our understanding of the naive estimator, we expect the second and third scenarios to have the same distribution and thus the same power.  We expect the same of the fourth and fifth scenarios.  In Figure~3, we see that these two expectations were met; namely, two pairs of power curves are effectively coincident.  Large gains in power are evident as we increase the range across the endpoints from 200g to 400g, whereby the maximum power for $\delta_{alt}=1.3$ is under 50\% in the former case and nearly 90\% in the latter.  Incremental gains, especially for more moderate  $\delta_{alt}$, are apparent when the range between weights grows to 600g in the three last scenarios.
While there is some gain from a four condition scenario, we contend that the bulk of the power gains are from extending the range of weight conditions.\footnote{Extending the range of weights without keeping the average distance between weights small may introduce complications in the clinical setting, so even though power is identical in the fourth and fifth scenarios, the fifth may be more realistic to execute.}

These various designs shed some light on what can be achieved in terms of power to detect non-normal behavior well-ahead of the actual clinical assessment.  Further, the clinician can weigh the cost of false negative errors against the burden to the patient in the assessment.

\begin{figure}[ht]
\begin{center}
%\framebox{\includegraphics[width=2in]{normL.png}}
\includegraphics[width=3.0in]{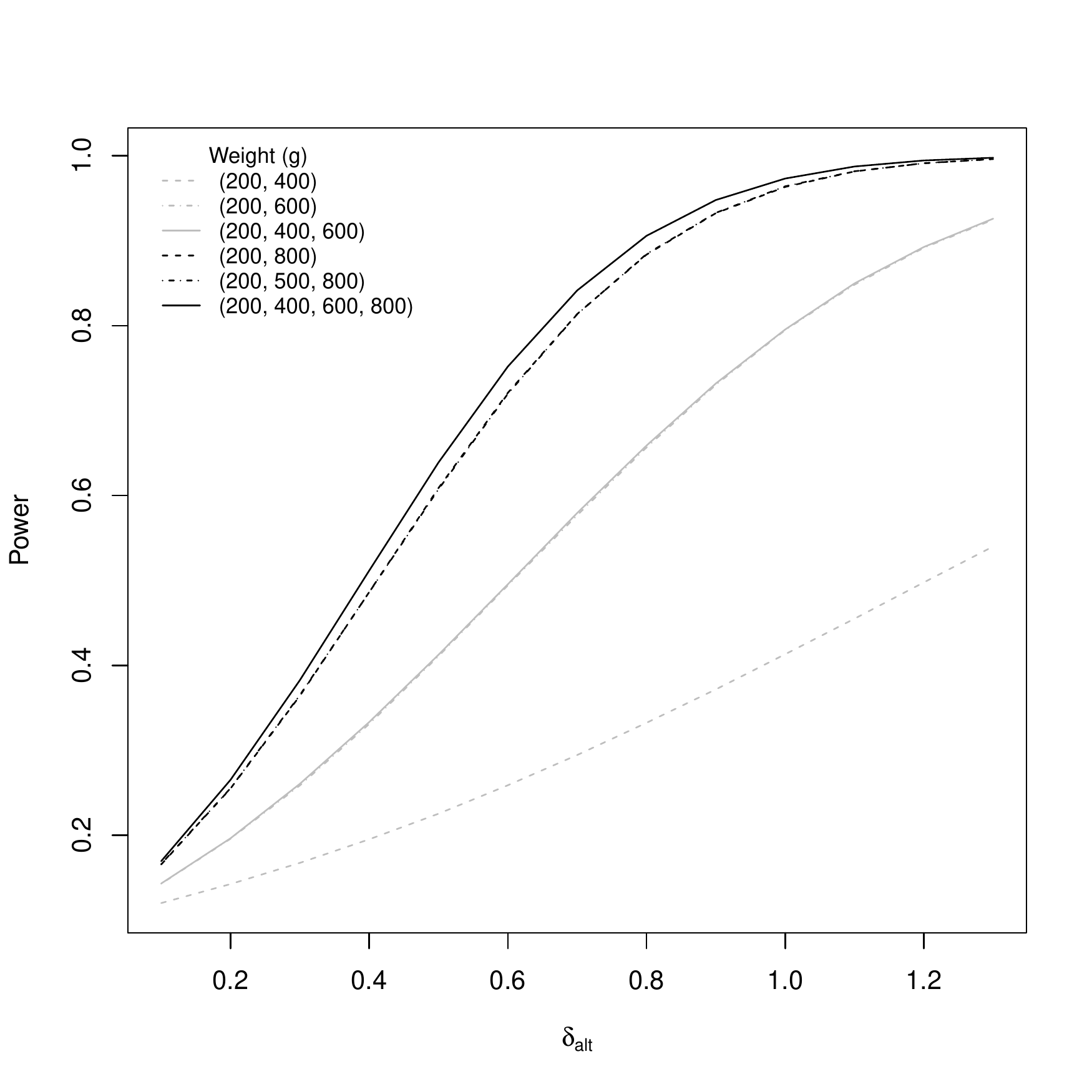}
\caption{Comparison of power for six test scenarios.  See text for details.  The power is calculated based on 100 realizations of $Dist(B*)$ under the null based on the proposed method.}
\end{center}
\end{figure}\label{power}

\section{An example of real data analysis}
In this section, we applied the proposed method to examine a group of patients with stroke from a single-subject design perspective. We are interested in understanding each patient's status as compared to a small group of healthy subjects in terms of their ability to predict the fingertip forces to object weight as measured by the scaling factor for PLFR, in two different experiments. The data was collected using protocols approved by the institutional review board, and informed consent was obtained from all subjects according to the Declaration of Helsinki. The data is described next.

\begin{description}
\item[Training Data Set] The training data set consists of data from 10 healthy subjects, each lifting 10 weights, ranging from 250 grams to 750 grams, 50 grams apart. The order of the weights is randomized to avoid an ordering effects. Each subject lifts each weight 6 times after one practice trial to learn the weight of the object.
\item[Test subject] There are 22 patients with stroke who are at various stages of post-stroke recovery. Each of the patients participate in two experiments to assess their fingertip force coordination with the affected hand.
\begin{description}
\item[Experiment One] Lift weights with affected hand, at 550 and 800 grams each.
\item[Experiment Two] Lift weights with affected hand following a practice lift with unaffected hand, 350 and 600 grams.
\end{description}
Each subject lifts each weight for 4 consecutive times after one practice lift. To avoid an ordering effect, the experimental conditions and weights within each condition are randomly assigned to the subjects.
\end{description}

The two experimental conditions one and two are intended to test whether the subjects show evidence of predictive control of fingertip forces to object weight during a grasp and lift task. Some patients suffer from sensory impairment in the affected hand, hence they may not be able to learn the weight of the object through practice using the affected hand alone only (experimental condition one), instead, such information may be learned by practicing with unaffected hand first (experimental condition two)\cite{Raghavan2006}. Ultimately, clinicians and researchers need this information to decide whether the patients should practice the grasping task with the affected hand alone, or incorporate the unaffected hand into practice protocols.

Using the data from test subjects, we first fit a linear hierarchical model to test the effects of experimental condition two as compared to condition one. At level 10\%, we found that the results of experimental condition two were significantly better in terms of the scaling factors. However these results from the entire group are not particularly useful when the clinician needs to make a decision and a recommendation of a practice protocol for any single patient during the course of their rehabilitation.

Using the proposed method, we can assess each subject separately under each experimental condition. Table~\ref{real_data} reports the results. Since most stroke rehabilitation treatment protocols are non-invasive and low-risk, we choose to tolerate a higher false positive rate, and set the level of the test to be at 10\%. The assessment results are therefore controlled at an expected 10\% false positive rate; the $p$-value and the (post hoc) power\footnote{This is the power one has to identify an effect of size equal to or greater than that which was actually attained.} of each assessment is also estimated and reported. This table reflects all the information available to the clinician.

Based on the test results, setting $\alpha=0.10$, we find that out of 18 subjects who completed both experimental conditions, six subjects switched status from ``ABNORMAL'' under condition one to "NORMAL" under condition two, one subject remained ``ABNORMAL'' in both conditions. Nine subjects remained ``NORMAL" and two switched from ``NORMAL'' to ``ABNORMAL''.
Clinicians can thus use these results to design customized training protocols for each patient.
Moreover, among those who receive an initial ``NORMAL'' assessment in experimental condition one, the information of post-hoc power and $p$-value can further inform clinicians about how effective the test is in detecting ``ABNORMAL'' status given the observed effect size, and the minimal false positive rate they have to accept if they choose to switch a ``NORMAL'' patient to the ``ABNORMAL'' status. Combining the information provided by the observed scaling factor, the power of the test, the $p$-value of the test and other patient-specific conditions, a clinician can make an informed choice to assign a particular patient to an appropriate training paradigm.

%**FILL** is this still relevant???
%{\bf For Table 2, we can probably present a confidence interval for $\delta$ by inverting the null distribution to get a range of variability}

\begin{table}
%\begin{adjustwidth}{-1in}{0in}
\tiny{
\begin{tabular}{c|ccccc|ccccc}
\hline
 & \multicolumn{5}{c}{Experiment One} & \multicolumn{5}{c}{Experiment Two} \\
 \hline
Subject & Assessment & Scaling Factor & $\delta$ & $p$-value & power & Assessment & Scaling Factor & $\delta$ & $p$-value & power \\
 1 & NORMAL    &    \phantom{-}1.989 &  -0.563  & 0.735 & 0.027  &  NORMAL   &  1.925 & -0.499  & 0.723 & 0.031   \\
 2 & NORMAL    &    \phantom{-}0.976 &   \phantom{-}0.449  & 0.282 & 0.221  &  NORMAL   &  1.664 & -0.238  & 0.603 & 0.056   \\
 3 & ABNORMAL  &    \phantom{-}0.135 &   \phantom{-}1.291  & 0.064 & 0.588  &  ABNORMAL &  0.278 &  \phantom{-}1.148  & 0.091 & 0.515      \\
 4 & NORMAL    &    \phantom{-}0.395 &   \phantom{-}1.031  & 0.105 & 0.469  &  N/A       &     -- &     --  &    --   \\
 5 & ABNORMAL  &   -0.330 &   \phantom{-}1.756  & 0.020 & 0.779  &  N/A       &     -- &     --  &    --   \\
 6 & NORMAL    &    \phantom{-}0.633 &   \phantom{-}0.793  & 0.165 & 0.360  &  NORMAL   &  0.683 &  \phantom{-}0.742  & 0.187 & 0.334 \\
 7 & NORMAL    &    \phantom{-}1.006 &   \phantom{-}0.420  & 0.293 & 0.209  &  N/A       &     -- &     --  &    --   \\
 8 & ABNORMAL  &   \phantom{-}0.172 &   \phantom{-}1.254  & 0.066 & 0.572  &  NORMAL   &  1.015 &  \phantom{-}0.410  & 0.314 & 0.200  \\
 9 & ABNORMAL  &   -1.293 &   \phantom{-}2.719  & 0.001 & 0.970  &  NORMAL   &  1.687 & -0.261  & 0.614 & 0.053  \\
10 & ABNORMAL  &   -0.158 &   \phantom{-}1.583  & 0.030 & 0.713  &  NORMAL   &  1.627 & -0.201  & 0.589 & 0.061   \\
11 & NORMAL    &    \phantom{-}1.308 &   \phantom{-}0.118  & 0.428 & 0.122  &  NORMAL   &  1.308 &  \phantom{-}0.117  & 0.445 & 0.121  \\
12 & NORMAL    &    \phantom{-}0.801 &   \phantom{-}0.625  & 0.220 & 0.483  &  ABNORMAL &  0.130 &  \phantom{-}1.296  & 0.064 & 0.581      \\
13 & ABNORMAL  &    \phantom{-}0.084 &   \phantom{-}1.342  & 0.059 & 0.613  &  NORMAL   &  1.849 & -0.423  & 0.691 & 0.036  \\
14 & ABNORMAL  &   -0.455 &   \phantom{-}1.881  & 0.013 & 0.815  &  NORMAL   &  2.632 & -1.207  & 0.928 & 0.004   \\
15 & NORMAL    &    \phantom{-}1.006 &   \phantom{-}0.420  & 0.293 & 0.209 &  NORMAL   &  1.609 & -0.183  & 0.580 & 0.065   \\
16 & NORMAL    &    \phantom{-}0.945 &   \phantom{-}0.481  & 0.269 & 0.232  &  NORMAL   &  1.715 & -0.290  & 0.628 & 0.050   \\
17 & NORMAL    &    \phantom{-}1.041 &   \phantom{-}0.385  & 0.312 & 0.199 &  NORMAL   &  1.983 & -0.557  & 0.745 & 0.025   \\
18 & NORMAL    &    \phantom{-}0.911 &   \phantom{-}0.515  & 0.258 & 0.245  &  ABNORMAL &  0.044 &  1.382  & 0.053 & 0.619      \\
19 & ABNORMAL  &    \phantom{-}0.268 &   \phantom{-}1.158  & 0.082 & 0.525  &  NORMAL   &  0.765 &  \phantom{-}0.661  & 0.211 & 0.298  \\
20 & NORMAL    &    \phantom{-}0.865 &   \phantom{-}0.561  & 0.243 & 0.259 &  NORMAL   &  1.818 & -0.393  & 0.679 & 0.039   \\
21 & NORMAL    &    \phantom{-}0.852 &   \phantom{-}0.573  & 0.239 & 0.263  &  N/A       &     -- &     --  &    --   \\
22 & NORMAL    &    \phantom{-}1.347 &   \phantom{-}0.079  & 0.447 & 0.115  &  NORMAL   &  1.370 &  \phantom{-}0.056  & 0.475 & 0.110  \\
\hline
\end{tabular}
}
%\end{adjustwidth}
\caption{The Physician's Chart. This table provides the information that will help physician to make an informed assessment about test subject's status. For each subject, the assessment decision is either ``Normal'' or ``Abnormal''. Since the true status of the subject is unknown, if the assessment is ``Normal'', we report the false negative rate (1-power) associated with this decision; if the assessment is ``abnormal'', we report the false positive rate (the level of the test) associated with this decision. Since the level of the test is set to be 10\%, the false positive rate associated with the positive decision ``abnormal'' is 10\%. Alternatively, we can report ``power'' for all the negative decision (``Normal'' cases).}
\end{table}\label{real_data}

%%% save the power analysis for future
%We can see that when the deviation from the null is small, small $\delta$ values, the false negative rate is high in all methods under scenario one. Since there is only one subject in the test set, the power of the test is low at the region near the null value. However, what is encouraging is that under scenario two, when the subject is asked to lift three weights rather than two, the information about the scaling factor increases quite a bit. The false negative rate is much smaller than under scenario one.  The power rate of different methods using the critical value derived from $g(\delta |\delta_0=0)$ is plotted in Figure Three. We can see that under scenario two, using the proposed method, and $\bar{\delta}$, the false positive rate is reasonably controlled when the null hypothesis is true (FPR=0.06), and the power increases quickly as $\delta_a$ deviates from 0. At $\delta_a=0.7$, the power of rejecting the null is about 0.8.  Clinically, this means that if a subject only scales about half of what's expected, we have 80\% chance of detecting the abnormality.

%The power plot is shown

%\begin{figure}[ht]
%\begin{center}
%\includegraphics[width=3in]{power.pdf}
%\caption{The Power plot for the Test A, B, C and D }
%\end{center}
%\end{figure}\label{power}

%need power plot to be sooner in text...

%I THINK THIS HAS BEEN DONE:
%%[**FILL**] Discussion of the clinical example goes here.

\section{Discussion}

%In this paper, we proposed a practical solution to the statistical testing problem regarding single-case design. In particular, we address an important clinical question: does the test patient behave the same as one from the healthy population? This question can not be answered using the traditional single subject design in which only the test subject information is used.\footnote{Comparison to population-based estimates do not take account the test conditions experimental design.} Borrowing the concept of training and test sets in machine learning, we propose using the Bayesian posterior predictive draws of the training subject data referenced or generated from the test subject design.  This yields a template null distribution of a test statistic for the purpose of inference prior to actual testing of new subjects. The performance of this template distribution can also be studied ahead of time. It can be used in a clinical situation and physicians can directly compare the quantity of interest to this distribution to make inference at any desired level. The simulation studies have shown that the proposed test performs satisfactorily when compared with its counterpart test in which the true sampling distribution of the test statistic is given or known. Moreover, we are able to provide an estimate of the error rate and its confidence interval given a single training data set, which can further inform physicians about the reliability of the test results based on the given template/experimental design.

In this paper, we have proposed a practical solution to the statistical testing problem in a single-subject design. This is important because in practice, clinicians have to make decisions for individual subjects, rather than for groups of subjects as in most randomized-controlled trials. In populations that are characterized by large amounts of within and between subject variability such as in rehabilitation medicine, the traditional evidence-based approaches that address only average group differences are not sufficient. A robust single subject design can help the clinician with critical individualized decision making. Borrowing from the concept of training and test sets in machine learning, we use  the Bayesian posterior predictive draws of the training subject data referenced or generated from the test subject design.  This yields a template null distribution of a test statistic for the purpose of inference prior to actual testing of new subjects. The performance of this template distribution can also be studied ahead of time. It can be used in a clinical situation and providers can directly compare the quantity of interest to this distribution to make an inference at any desired level. The simulation studies have shown that the proposed test performs satisfactorily when compared with its counterpart test in which the true sampling distribution of the test statistic is given or known. Moreover, we are able to provide an estimate of the error rate and its confidence interval given a single training data set, which can further inform clinicians about the reliability of the test results based on the given template/experimental design.

Compared to the traditional single-subject design approaches, our proposed method has the following advantages:
\begin{enumerate}
\item  Our proposed method can be easily adapted to test a range of quantities of interest. In this paper, we study an example using a statistic based on simple mean difference, but clinicians may be interested in using the between trial variability as a measure of performance, for example. Using the algorithm in section~\ref{algorithm}, the template distribution of the between-trial variability can be conveniently produced based on the between-trial variability of $\tilde{Y}_{ijt}$.
\item Our proposed test is based on a small sample of training subject data. In general, we expect that the training subject data is a random sample and the experiment is done in a well-defined laboratory setting. Albeit a small sample size (for example $N=10$ subjects), a more complex design that better informs the between-subject and within-subject variability can be used. For example, in this paper, we consider a training subject design using 10 weights and more trial replications.  In contrast, we allow the test subject design to be simpler so that it is feasible for patients in a clinical setting.
\item Another advantage of the proposed method is that it can be used to inform single-subject design. For example, the power analysis suggested that the most effective way of improving the power of the test is to diversify the conditions (increasing the distance in weight between two conditions) rather than increasing the number of repeated trials; given a fixed weight range,  two-weight design and three-weight design are equivalent. Moreover, since the error rates associated with different test designs can be approximately calculated, researchers and clinicians can determine, ahead of time, which experimental design for the test subject optimizes the error rates and power among all feasible options.
\item Lastly, in a tele-medicine situation, our proposed method will most effectively allow the outcomes from a basic research laboratory to be quickly applied in a clinical situation, where clinicians can download a template distribution provided from the research labs and upload clinical data to enable effective evidence-based decision making.

%when the physicians are able to download a template distribution provided from the research labs and upload clinical data to the lab, our proposed method will most effectively allow the outcomes in the basic research laboratory to be quickly applied in a clinical situation.  In addition, it allows for the rapid integration of newly collected clinical data for the purpose of basic research.
\end{enumerate}

%\bibliography{rehab}

\end{document}